\documentclass{article}

\usepackage{arxiv}
\usepackage[utf8]{inputenc}

\usepackage{tikz,pgfplots}
\usetikzlibrary{shadows.blur}
\usetikzlibrary{decorations.shapes}
\usetikzlibrary{decorations.pathmorphing}
\usepgfplotslibrary{fillbetween}
\pgfplotsset{compat=newest}
\usepackage[
    natbib=true,
    style=numeric,
    sorting=none
]{biblatex}
\usepackage{biblatex}
\usepackage[english]{babel}
\usepackage{csquotes}
\usepackage[T1]{fontenc}    
\usepackage{hyperref}
\hypersetup{
    colorlinks=true,
    linkcolor=red,
    filecolor=magenta,      
    urlcolor=blue,
    pdftitle={Overleaf Example},
    pdfpagemode=FullScreen,
    }
\usepackage{authblk}
\usepackage{booktabs}       
\usepackage{amsfonts}       
\usepackage{nicefrac}       
\usepackage{microtype}      
\usepackage{lipsum}		
\usepackage{amsmath}
\usepackage{graphicx}
\usepackage{doi}
\usepackage{graphicx}
\usepackage{dcolumn}
\usepackage{bm}
\usepackage{amssymb}

\title{Force-Free Electromagnetic Configurations in Arbitrary Geometries}

\author[1,\thanks{radhikari@troy.edu}]{Rakshak Adhikari }

\affil[1]{Center for Relativity and Cosmology\\ Troy University, Troy, AL 36082}

\renewcommand{\shorttitle}{{FFE in Arbitrary Geometries}}

\begin{document}
\maketitle
\begin{abstract}

The dynamics of highly magnetized plasmas in extreme astrophysical environments are effectively modeled by Force-Free Electrodynamics (FFE), a framework essential for studying objects like neutron stars and accreting black holes. The inherently nonlinear nature of the FFE equations makes finding exact solutions a challenging task. This paper explores an innovative approach to solving these equations by foliating spacetime into two-dimensional surfaces, specifically tailored to the geometry of electromagnetic fields. The foliation approach exploits the fact that the kernel of the force-free field defines an involutive distribution, naturally lending itself to a geometric decomposition. This method has previously been applied with great success in Kerr and FLRW spacetimes. By extending this formalism, we develop new exact solutions to the FFE equations in several distinct spacetimes, including cases where the metric remains partially undetermined. Additionally, we present a novel class of solutions that smoothly transition between magnetically dominated and electrically dominated regimes over time. Finally, we construct a pair of vacuum degenerate fields in arbitrary axisymmetric spacetimes, further demonstrating the utility of the foliation approach. These results offer a new perspective on the dynamical evolution of electromagnetic fields and their geometric underpinnings in various spacetime geometries, broadening the applicability of the framework for understanding the dynamics of highly magnetized plasmas in general relativistic astrophysical contexts.
\end{abstract}


 \section*{Introduction}

Force-free electrodynamics (FFE),the low-inertia limit of magnetohydrodynamics (MHD), describes the behavior of highly magnetized plasmas in regimes where the energy density of the electromagnetic field dominates over matter contributions. This framework is essential for understanding astrophysical environments such as the magnetospheres of neutron stars and accreting black holes.

Soon after the discovery of pulsars, Goldreich and Julian \cite{Goldreich:1969sb} presented their model of a pulsar as a highly conductive, magnetized neutron star with its spin axis aligned with its magnetic dipole. They demonstrated that such an object would generate an intense electric field, leading to the formation of a force-free magnetosphere around it. Later, Blandford and Znajek \cite{BZ77} showed that a force-free plasma surrounding a slowly rotating black hole can extract rotational energy from the black hole’s spin in the form of Poynting flux.

Despite their utility, the equations of force-free electrodynamics present a significant challenge due to their inherent nonlinearity, often precluding analytical solutions. As a result, studies of FFE are largely reliant on numerical simulations \cite{Talbot:2020zkb,Koide:2025bhs,Mahlmann:2024gui, Urban:2023cfk}, as well as perturbative approaches \cite{Camilloni:2022kmx} following in the footsteps of Blandford and Znajek. However, analytical efforts remain crucial for guiding and validating numerical results, as well as for providing fundamental theoretical insights and identifying key physical parameters. Early theoretical investigations into FFE, such as those in \cite{Uchida1,Uchida2}, laid the groundwork for analytical studies. Subsequently, specific classes of exact solutions were discovered under additional simplifying assumptions \cite{Menon:2005mg,Menon15, Brennan:2013kea}. A major advancement came with the work of Gralla and Jacobson \cite{Gralla:2014yja}, who reformulated FFE within the framework of exterior calculus, offering a more geometric and covariant approach. This was followed by the first systematic study of foliations and force-free fields in \cite{Compere:2016xwa}.

Observational evidence strongly supports the role of force-free electrodynamics in high-energy astrophysics. The correlation between black hole spin and relativistic jet power\cite{Steiner:2012ap} aligns with the Blandford-Znajek mechanism, which describes the energy extraction from rotating black holes via force-free fields. Additionally, force-free models have been proposed \cite{Hada:2015okc} to explain the limb-brightened structure of the relativistic jet in M87, and Very Long Baseline Interferometry (VLBI) observations \cite{park2021jet} have shown consistency with steady, axisymmetric force-free solutions. These observations motivate the continued search for exact analytical solutions to deepen our understanding of these astrophysical processes.

A recent breakthrough in force-free electrodynamics has emerged through the study of spacetime foliations. In a series of papers \cite{Menon:2020hdk, Menon:2020ivu, Menon:2020npo} beginning in 2020, Menon introduced a novel geometric approach that exploits the properties of foliations to construct exact force-free fields. By 2023, this formalism successfully generated several exact solutions in Kerr spacetime \cite{Adhikari:2023hhk} as well as the FLRW spacetime \cite{Adhikari:2024hre}, demonstrating its potential for systematically solving force-free equations in diverse spacetime settings.

This paper expands on this foliation-based approach, demonstrating its power in constructing both null and non-null force-free solutions across different spacetimes. Specifically, we show how the formalism applies to partially determined metric components, allowing for the construction of force-free fields in arbitrary spherically symmetric and axisymmetric spacetimes. By pedagogically illustrating the methodology, we provide new insights into the general structure of force-free solutions.

The significance of this work extends beyond mathematical formalism. Although the Kerr solution is widely used to model rotating black holes, its validity as the definitive description of astrophysical black holes remains an active area of inquiry. Recent challenges \cite{Carballo-Rubio:2024dca} to this idea suggest that the interiors of rotating black holes may be more complex than previously thought, motivating the exploration of alternative models. These alternatives include models that incorporate the regularity at the horizon \cite{Li:2023nmy} or generalize the Kerr metric by adding new deviation parameters \cite{Johannsen:2012mu, Pei:2016kka}. Exploring force-free fields in a broader range of spacetimes may contribute to the identification of electromagnetic field characteristics that are specific to certain models versus those that are more universally observed.

\section*{Equations of Force-Free Electrodynamics}
Maxwell's equation in an arbitrary spacetime is given by
\begin{equation}
 d  F = 0\;, \;\;\;{\rm and}
\;\;\;*d*F = j\;.
\label{inhomMaxform}
\end{equation}
Here $F$ is the Maxwell field tensor, $*$, the Hodge-Star operator, $d$, the exterior derivative on forms, and $j$ denotes the current density dual vector. Throughout this work, we adopt the exterior calculus formulation of force-free electrodynamics laid out in \cite{Gralla:2014yja}.
Force-free electrodynamics is defined by the constraint $F_{\mu\nu} j^\nu = 0$.
The Maxwell Field tensor $F$ is said to be magnetically dominated whenever $F^2 >0$, $F$ is electrically dominated whenever $F^2 <0$, and finally a force-free electromagnetic field $F$ is null whenever $F^2 =0$. As usual $F^2 =F_{\mu\nu}F^{\mu\nu}$.

The kernel of $F$ is the set of all tangent vectors that annihilate $F$. In force-free electrodynamics, the current density $j$ is always such a vector field. It is well known that the kernel of $F$ forms the tangent space to a two-dimensional manifold.

\subsection*{Geometric Framework for Force-Free Electrodynamics}

Force-free electromagnetic fields are typically described by a degenerate 2-form field strength $F$, which is defined as

\begin{equation}
    F = \alpha \wedge \beta,
\end{equation}

where $\alpha$ and $\beta$ are 1-forms. The kernel of $F$, denoted $\ker F$, represents the two-dimensional subspace spanned by forms $u$ and $v$ that annihilate $F$. This kernel satisfies the condition

\begin{equation}
    \ker(F) = \{\textrm{span}(u,v) : \alpha(v) = \alpha(u) = \beta(v) = \beta(w) = 0\}.
\end{equation}

The leaves of this foliation, denoted ${\cal F}_a$, are called field sheets or flux surfaces, and they satisfy the properties

\[
    {\cal F}_a \cap {\cal F}_b = 0 \quad \text{for} \quad a \neq b, \quad \bigcup_{a \in A} {\cal F}_a = {\cal M}.
\]

These foliation leaves partition the spacetime manifold into a set of distinct submanifolds. The general theory of FFE is categorized into magnetically and electrically dominated solutions, which are geometrically similar but differ in the details of their behavior. It is important to note that the magnetically dominated solutions are well-posed, whereas the electrically dominated solutions may lead to divergences in the field tensor as they approach the null limit \cite{Menon:2020npo}.

\subsection*{Non-null Force-Free Fields}

Around any point $p$ on a manifold ${\cal M}$, we can find an inertial frame $\{e_\mu\}$, where $\mu = 0, 1, 2, 3$. If the Lie bracket of $e_0$ and $e_1$ satisfies 
\[
[e_0, e_1] \in \text{span}\{e_0, e_1\},
\]
then we have a foliation of the manifold ${\cal M}$ into submanifolds, where the leaves of the foliation are spanned by $e_0$ and $e_1$.

Given such vector fields $e_0$ and $e_1$ tangent to the leaves of the foliation, the shape tensor or the second fundamental form $\Pi$ is defined by 
\[
\Pi(e_0, e_1) = (\nabla_{e_0} e_1)^\perp,
\]
where $V, W$ are vector fields tangent to the leaves, and $\perp$ denotes the component normal to the surface.

The mean curvature field $H$ can then be defined as 
\[
H = \frac{1}{2} \left[- \Pi(e_0, e_0) + \Pi(e_1, e_1)\right].
\]

Similarly, even though the remaining two vector fields $e_2$ and $e_3$ may not define a submanifold, we can define the dual mean curvature field $\tilde{H}$ as 
\[
\tilde{H} = \frac{1}{2} \left[\Pi(e_2, e_2) + \Pi(e_3, e_3)\right].
\]

\begin{figure}[htbp!]
    \centering
\begin{tikzpicture}[scale=1]
\begin{axis}[xmin=-4, xmax=7,   ymin=-5,   ymax=7,zmin=-50,zmax=300 ,
    title=,
    hide axis
    , colormap/viridis,opacity=0.8];
\addplot3[
    mesh,samples=11,domain=-5:5,]{-((x)^2+(y)^2)};
\addlegendentry{$[e_0,e_1]\in \textrm{span}(e_0,e_1)$};

\tikzset{myrectstyle/.style={dashed,fill=gray!70,opacity=0.5}}
    \addplot3 [myrectstyle] ({-4},{1},{90})  -- ({-2},{3},{110}) -- ({-2},{3},{240})-- ({-4},{1},{220}) -- cycle;
\node at (5,5,280) [below]{$ d(H+\Tilde{H})=0$};
\node at (5,5.5,220) [below]{$ F=u\:e_2^\flat\wedge e_3^\flat$};
\draw [thick,dashed] (-3,2,-10) -- (-3,2,100);
\draw [->,thick] (3,1,0) -- (3,1,142) node [above] {$ H$};
\draw [->,thick] (-3,2,-10) -- (-3,-1,-5) node [below] {$e_1$};
\draw [->,thick] (-3,2,-10) -- (-1,2,-3) node [right] {$e_0$};

\draw [->,thick] (-3,2,145) -- (-1.9,2,230) node [right] {$e_3$};
\draw [->,thick] (-3,2,145) -- (-2.4,-2.2,230) node [left] {$e_2$};
\draw [->,thick] (-3,2,145) -- (-0.5,3.5,150)node [above] {$\Tilde{H}$};
\end{axis}
\end{tikzpicture}
    \caption{Time-like Foliation Generating a Magnetically Dominated Force-free Field}
\end{figure}
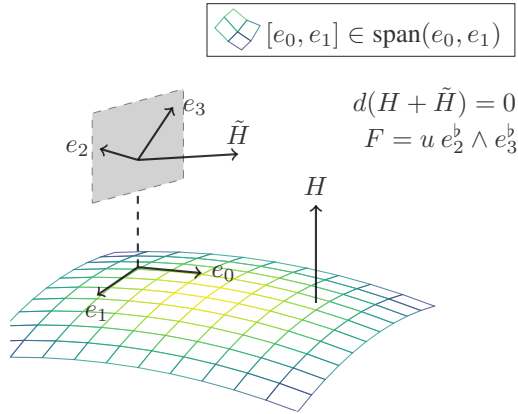

It was shown that for an involutive distribution defined by the vector fields $e_0$ and $e_1$, a unique force-free field exists, given by 
\[
F = u \; e_2^\flat \wedge e_3^\flat,
\]
if and only if 
\[
d(H + \tilde{H}) = 0,
\]
where $H$ and $\tilde{H}$ denote the mean and dual mean curvature fields, respectively. 

Here, $u$ is given by 
\[
d(\ln u) = 2(H + \tilde{H})^\flat.
\]

\subsection*{Electric Dominated Force-Free Fields}

An analogous result applies to electrically dominated force-free fields. In this case, the kernel of the field is spanned by $e_2$ and $e_3$, and the force-free field is given by

\[
    F = u \, e_0^\flat \wedge e_1^\flat.
\]

The expression for the function $u$ remains the same as in the magnetically dominated case, and it satisfies the equation

\[
    d (\ln u) = 2 (H + \tilde{H})^\flat.
\]

\begin{figure}[htbp!]
\centering
\begin{tikzpicture}[scale=1]
\begin{axis}[xmin=-4, xmax=7,   ymin=-5,   ymax=7,zmin=-50,zmax=300 ,
    title= ,
    hide axis
    , colormap/viridis,opacity=0.8];
\addplot3[
    mesh,samples=11,domain=-5:5,]{-((x)^2+(y)^2)};
\addlegendentry{$[e_2,e_3]\in \textrm{span}(e_2,e_3)$};

\node at (5,5.5,220) [below]{$ F=u\:e_0^\flat\wedge e_1^\flat$};
\tikzset{myrectstyle/.style={dashed,fill=gray!70,opacity=0.5}}
    \addplot3 [myrectstyle] ({-4},{1},{90})  -- ({-2},{3},{110}) -- ({-2},{3},{240})-- ({-4},{1},{220}) -- cycle;
\node at (5,5,280) [below]{$ d(H+\Tilde{H})=0$};
\draw [thick,dashed] (-3,2,-10) -- (-3,2,100);
\draw [->,thick] (3,1,0) -- (3,1,142) node [above] {$ H$};
\draw [->,thick] (-3,2,-10) -- (-3,-1,-5) node [below] {$e_2$};
\draw [->,thick] (-3,2,-10) -- (-1,2,-3) node [right] {$e_3$};

\draw [->,thick] (-3,2,145) -- (-1.9,2,230) node [right] {$e_0$};
\draw [->,thick] (-3,2,145) -- (-2.4,-2.2,230) node [left] {$e_1$};
\draw [->,thick] (-3,2,145) -- (-0.5,3.5,150)node [above] {$\Tilde{H}$};
\end{axis}
\end{tikzpicture}
  \caption{Space-like Foliation Generating an Electrically Dominated Force-free Field}
\end{figure}
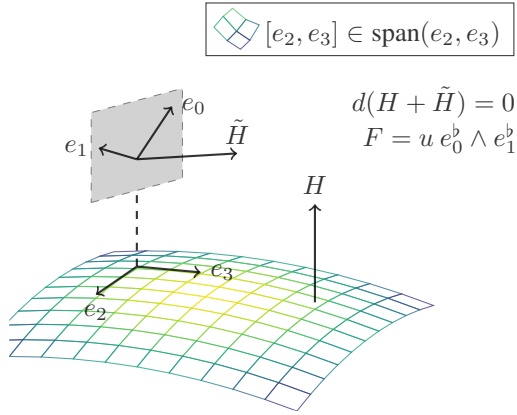

\section*{Null Foliations}
For the study of null foliations we start with a null pregeodesic congruence i.e a null vector l such that $l^\mu \nabla_\mu l^nu\propto l^\nu$. Given such a vector field, we can construct a null tetrad $e_i$ whose basis are  s, l, $\alpha^\#$ and n, such that

\begin{equation}
g(e_i,e_j)=\left[\begin{array}{cccc}
1 & 0 & 0 & 0 
\\
 0 & 0 & 0 & -1 
\\
 0 & 0 & 1 & 0 
\\
 0 & -1 & 0 & 0 
\end{array}\right],
\end{equation}

and $(l,s)$ form an involutive distribution. Such a tetrad is called a null foliation adapted frame.\footnote{We can find specialized coordinate charts $\big(U_p, \phi_p = (x^1, \dots, x^4)\big)$ that align with the foliation. In these charts, the leaves of the foliation correspond to surfaces where $x^3$ and $x^4$ remain constant.}

We can calculate the null mean curvature $\theta$ for the congruence generated by l,

\begin{equation}
    \theta = \frac{1}{2} \left[ g(\nabla_s l, s) + g(\nabla_{\alpha^\sharp} l, \alpha^\sharp) \right].
\end{equation}

The frame defined by the null tetrad is said to admit an equipartition of null mean curvature with respect to the null pregeodesic vector field l if
$$g(\nabla_s l, s) =g(\nabla_{\alpha^\sharp} l, \alpha^\sharp).$$

It was shown that a null foliation adapted frame gives rise to a unique class of null and force-free field if it admits an equipartition of the null mean curvature with respect to the null pregeodesic vector field l.
The null force-free field is then given by 
\begin{equation}
F = (u \cdot \kappa) \; \alpha \wedge l^\flat.
\end{equation}
In the null foliation adapted chart $(x^1, x^2, x^3, x^4)$ where field sheets are given by the condition $x^1, x^2 = {\rm const}$,
\begin{equation}
  \kappa=(\alpha_3 \;l^\flat_4-\alpha_4\; l^\flat_3)^{-1}\;,
  \label{kappadef}
\end{equation}
where 
\begin{equation}
  \left(
           \begin{array}{c}
             \alpha \\
             l^\flat \\
           \end{array}
         \right)=\left(
    \begin{array}{cc}
      \alpha_3 & \alpha_4 \\
      l^\flat_3 & l^\flat_4  \\
    \end{array}
  \right)\left(
           \begin{array}{c}
             dx^3 \\
             dx^4 \\
           \end{array}
         \right)\;. 
         \label{chart2frame}
\end{equation}
Here $u = u(x^3, x^4)$ is any smooth function of $x^3$ and $ x^4$, and this means that null force-free fields always exist with a two-parameter freedom that could be constrained to match the physical conditions or regularity conditions at hand.
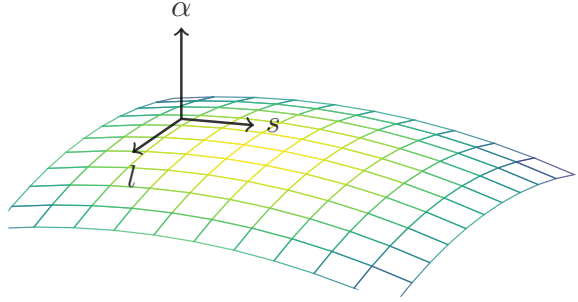
\begin{figure}[htbp!]
\centering
\begin{tikzpicture}[scale=1.2]
\begin{axis}[xmin=-4.6, xmax=7,   ymin=-5,   ymax=7,zmin=-50,zmax=300 ,
    title= ,
    hide axis
    , colormap/viridis,opacity=0.8];
\addplot3[
    mesh,samples=11,domain=-5:6,]{-((x)^2+(y)^2)};
\addlegendentry{$[l,s]\in \textrm{span}(l,s)$};

\node at (4.5,4.5,260) [below]{$ F=(u\cdot \kappa)\:\alpha \wedge l^\flat$};

\draw [->,thick,] (-3,2,-10) -- (-3,2,100) node [above] {$\alpha$};

\draw [->,thick] (-3,2,-10) -- (-3,-1,-5) node [below] {$l$};
\draw [->,thick] (-3,2,-10) -- (-1,2,-3) node [right] {$s$};
\end{axis}
\end{tikzpicture}
  \caption{Null Foliation Adapted Frame}
\end{figure}

\section*{A non-null force-free field}
First we look at a non-null foliation in the Schwarzschild spacetime. The line element in the usual radial coordinates is given by

\begin{equation}
    ds^2 = -\left(1 - \dfrac{2M}{r}\right) dt^2 + \left(1 - \dfrac{2M}{r}\right)^{-1} dr^2 + r^2 d\theta^2 + r^2 \sin^2\theta \, d\varphi^2.
\end{equation}
We begin by considering the following canonical tetrad \footnote{These tetrads are the $a \rightarrow 0$ limit of the canonical tetrad given by O'neill \cite{ONeill2014}.}

\begin{subequations} \begin{eqnarray} e_0 &=& \frac{1}{r \sqrt{r^2 - 2Mr}}\: \partial_t \;, \\ e_1 &=& \frac{1}{r \sin \theta}\: \partial_\varphi \;, \\ e_2 &=& \sqrt{\frac{r^2 - 2Mr}{r^2}}\: \partial_r \;, \\ e_3 &=& \frac{1}{r}\: \partial_\theta \;. \end{eqnarray} \label{F2L1}
\end{subequations}

The tetrad is transformed as follows,

$$ \begin{pmatrix}
\bar e_0\\
\bar e_1\\
\bar e_2\\
\bar e_3\\
\end{pmatrix}= \begin{pmatrix}
A & 0 & -\sqrt{(A^2-1)} & 0 \\
0 & 0 & 0 & 1 \\
-\sqrt{(A^2-1)} & 0 & A & 0 \\
0 & 1 & 0 & 0
\end{pmatrix}
\begin{pmatrix} e_0\\
e_1 \\
e_2\\
e_3\\
\end{pmatrix}\;$$
where

$$A = \frac{1}{2}\left(f \cdot \exp\left(\frac{k t}{M}\right) + \frac{1}{f} \cdot \exp\left(-\frac{k t}{M}\right)\right).$$
Here f is an arbitrary function of $\theta$ and k is a constant. It is easy to check that $(\bar{e}_2,\bar{e}_3)$ forms an involutive pair. We can then calculate the mean curvature field H and its dual $\tilde{H}$ using
\begin{equation}
 2H= \left[-g(\nabla_{e_0} e_0, e_2)+ g(\nabla_{e_1} e_1, e_2) \right] e_2 + \left[-g(\nabla_{e_0} e_0, e_3)+ g(\nabla_{e_1} e_1, e_3) \right] e_3\;, 
    \label{Hdef}
\end{equation}
and
\begin{equation}
 2\tilde H= \left[-g(\nabla_{e_2} e_2, e_0)- g(\nabla_{e_3} e_3, e_0) \right] e_0 + \left[g(\nabla_{e_2} e_2, e_1)+ g(\nabla_{e_3} e_3, e_1) \right] e_1\;.  
   \label{Htildedef}
\end{equation}
We have 
\begin{equation}
    2(H^\flat+\tilde{H}^\flat)=\dfrac{k\:r^2+M^2-M\:r}{Mr(r-2M)}\:dr-\cot{\theta}\: d\theta
\end{equation}
It is easy to check that $d (H^\flat+\tilde{H}^\flat)=0$. Since all the requirements for the foliation are satisfied, our electrically dominated force-free field is given by
\begin{align}
    F_1&=u\:e_0^\flat\:\wedge e_1^\flat\\
    &= u\:(\sqrt{r^2-2Mr}\:A\:dt \wedge d\theta\:+\:\dfrac{r^2\:\sqrt{(A^2-1)}}{\sqrt{r^2-2Mr}}\:dr \wedge d\theta),
\end{align}

where $u=\dfrac{u_0(r-2M)^{2k-\frac{1}{2}}e^{\frac{kr}{M}}}{\sqrt{r}\:\sin{\theta}}.$

Here, $F_1^2=-2u^2$ is manifestly electrically dominated, and the 4-current density is given by

\begin{equation}
    J_1=\dfrac{u_0 (r-2M)^{2k+\frac{1}{2}} e^{\frac{k(r-t)}{M}} f'}
{2\sqrt{c(r^2-2Mr)} f^2 \sin\theta}\:\left( \frac{f^2 \:e^{\frac{2kt}{M}} -1}{r-2M} \:\partial_t 
- \dfrac{f^2 \:e^{\frac{2kt}{M}} +1}{r}\: \partial_r \right).
\end{equation}

Interestingly, we can have a vacuum degenerate field given by $F_2=c_0\:\sin{\theta}\: d\theta \wedge d\varphi$ such that $g(F_2,J_1)=0$ and $F_2^2=\dfrac{2c_0^2}{r^4}$.

So we can construct a new force-free field \( F_1' = F_1 + F_2 \) such that  
\[
{F_1'}^2 = \frac{1}{r^4} \left( -u_0^2 r^3 (-2M + r)^{4k - 1} \csc^2(\theta) \exp\left(\frac{2kr}{M}\right) + 2c_0^2 \right).
\]

The field is null when  
$$
2c_0^2 = u_0^2 r^3 (-2M + r)^{4k - 1} \csc^2(\theta) \exp\left(\frac{2kr}{M}\right).
$$ 
It is magnetically dominated (\( F^2 > 0 \)) if  
$$
2c_0^2 > u_0^2 r^3 (-2M + r)^{4k - 1} \csc^2(\theta) \exp\left(\frac{2kr}{M}\right),
$$ 
and electrically dominated (\( F^2 < 0 \)) if the inequality is reversed. Since this is a type-changing solution, the standard tetrad formalism we introduced will not suffice to describe the folitation that gives rise to this field. We refer the readers to  foliation adapted chart first described in \cite{Menon:2020npo} that can describe such solutions.

\section*{A null force-free field}
We demonstrate the search for null solutions in the case of an arbitrary spherically symmetric space-time, given by the following line element:

\begin{equation}
    ds^2 = -f \, dt^2 + h \, dr^2 + r^2 \left( d\theta^2 + \sin^2\theta \, d\varphi^2 \right),
\end{equation}
where \( f(r) \) and \( h(r) \) are arbitrary positive functions of the radial coordinate system.

We have the null geodesic congruence given by 

$$l=\dfrac{1}{\sqrt{f}}\:\partial_t+\dfrac{1}{\sqrt{f\:g}}\:\partial_r.$$

Similarly we have the following vector fields

$$\alpha^\sharp=\dfrac{1}{r}\:\partial_\theta \hspace{0.5cm}\text{and,}\hspace{0.5cm}s=\dfrac{1}{r\:\sin{\theta}}\:\partial_\varphi.$$

A simple calculation shows that the equipartition condition is satisfied and the null force-free field is given by

\begin{equation}
    F= u(\xi,\theta)\: (dt \wedge d\theta+\sqrt{\frac{h}{f}}\:dt \: \wedge \:d\theta),
\end{equation}
where $\xi=\int \sqrt{\frac{h}{f}}\:dr+t$.

\section*{A temporally type-changing solution in flat space}
In Minkowski space in Cartesian coordinates, we have
\begin{equation} 
    F_2 = \sqrt{f_3+ c_2} \: dt \wedge dx + c_1 (f_1 - f_2) \: dt \wedge dz + (f_1 + f_2) \: dx \wedge dz, 
\end{equation}
where  
\begin{equation}
    f_1 = e^{c_1 t + x}, \quad f_2 = e^{c_1 t - x}, \quad 
    f_3 = (1 - c_1^2)(f_1 - f_2)^2.
\end{equation}

The current 4-vector is given by  
\begin{equation}
    J_2= -\dfrac{1}{2}\:\dfrac{\partial_x\: f_3}{\sqrt{f_3+c_2}}\:\partial_t
    -\dfrac{1}{2}\:\dfrac{\partial_t \:f_3}{\sqrt{f_3+c_2}}\:\partial_x
    + (1-c_1^2)(f_2-f_1) \:\partial_z,
\end{equation}
and we have 
\begin{equation}
    F_2^2= 8\:e^{2c_1\:t}-2c_2.
\end{equation}

The field is electrically dominated when \( F_2^2 < 0 \), which occurs if \( 8 e^{2c_1 t} - 2c_2 < 0 \), which means that for sufficiently small \( t \), the electric component dominates provided \( c_2 > 0 \). The field is null at  
\begin{equation}
    t = \frac{1}{2c_1} \ln \left(\frac{c_2}{4}\right),
\end{equation}
if \( c_2 > 0 \), and becomes magnetically dominated (\( F_2^2 > 0 \)) for sufficiently large \( t \) when \( c_1 > 0 \), indicating an eventual dominance of the magnetic field over time.

Although this field solution lacks an immediate astrophysical interpretation, its significance lies in demonstrating a non-trivial temporal evolution within a force-free field. As shown in \cite{Adhikari:2024hre} and the previous section, spatial transitions in force-free fields have been previously demonstrated. Here, we establish that such transitions can also occur temporally. Specifically, this solution illustrates a smooth transition from an electrically dominated state, through a null field configuration, to a magnetically dominated state. This smooth, continuous type-change contrasts with the abrupt transitions often encountered in numerical simulations of neutron star and black hole magnetospheres, highlighting that force-free field theory allows for such gradual evolution without the formation of current sheets.

\section*{Degenerate vacuum fields in arbitrary axisymmetric spacetime}
The line element for an arbitrary axisymmetric spacetime as given in \cite{Chandrasekhar:1985kt} can be written as:

\begin{equation}
    ds^2 =- e^{\nu} dt^2 + e^{2\psi} (d\varphi - \omega dt)^2 + e^{2\nu_2} dr^2 + e^{2\nu_3} d\theta^2,
\end{equation}

where \( \nu \), \( \psi \), \( \nu_2 \), and \( \nu_3 \) are arbitrary functions of the coordinates.

\begin{subequations}
    \begin{align}
        e_0&=\sqrt{c^2 +e^{-\nu}}\:\partial_t+\:\left( e^{-\nu/2}\:\omega\:\sqrt{e^\nu c^2+1}+c\:e^{\nu/2-\psi}\right) \;\partial_\varphi\\
        e_1&= c\:\partial_t+\left(c\:\omega+e^{-\psi}\sqrt{e^\nu c^2+1}\right)\:\partial_\varphi\\
        e_2&= \:e^{-\mu_2}\:\partial_r\\
        e_3&=  \: e^{-\mu_3}\:\partial_\theta
    \end{align}
\end{subequations}

It turns out both pairs $(e_0,e_1)$ and $(e_2,e_3)$ are involutive. The sum of the mean curvature and its dual is closed and is given by:
\begin{equation}
    2(H+\Tilde{H})=-\dfrac{1}{2} \left[( \partial_r\nu+2\:\partial_r \psi)\:dr+(2\:\partial_\theta \:\psi+\partial_\theta \nu)\right]
\end{equation}

Here, u is given by
\begin{equation}
    u=u_0\:e^{-\psi-\nu/2}.
\end{equation}

Therefore, the electrically and magnetically dominated solutions are given by:
\begin{align}
    F_4&= u\: e_0^\flat \wedge e_1^\flat= u_0\:e^{-\nu/2-\psi+\mu_2+\mu_3} \:dr \wedge d\theta,\\
    F_5 &= u\: e_2^\flat \wedge e_3^\flat=u_0 \:dt \wedge d\phi. \nonumber
\end{align}

The current vanishes \footnote{If both pairs $(e_0,e_1)$ and $(e_2,e_3)$ describe involutive distributions, and if $d(H^\flat+\tilde{H^\flat})=0$, we have a set of complimentary vacuum degenerate fields that are magnetically and electrically dominated respectively.} for both the solutions so they are trivially force-free or vacuum degenerate i.e.
\begin{equation}
    J_4=J_5=0.
\end{equation}
And we have 
\begin{align}
    F_4^2&=-2\:u_0^2 \:e^{-\nu-2\psi},\\
    F_5^2&=2\:u_0^2 \:e^{-\nu-2\psi}.
\end{align}
In the special case of the Kerr metric, the metric functions take their familiar Boyer--Lindquist forms, and the electrically dominated solution reduces to
\begin{equation}
    F_4 =  \frac{u_0\:\rho^2}{\Delta\, \sin\theta}\, dr \wedge d\theta,
\end{equation}
where \(\rho^2 = r^2 + a^2 \cos^2\theta\) and \(\Delta = r^2 - 2 M r + a^2\). 

{In the Kerr case, solution $F_4$ is the vacuum limit of the solution presented \cite{Adhikari:2023hhk} whereas solution $F_5$ is the Hodge-star of the vacuum limit of the solution presented in \cite{Menon15}}. 

\section*{Conclusion}
This paper explored the construction of force-free electromagnetic configurations in various geometries using spacetime foliations. By extending the foliation-based approach, new exact solutions to the force-free electrodynamics equations were developed in diverse spacetimes, including cases with partially undetermined metrics. A novel class of solutions exhibiting smooth transitions between magnetically and electrically dominated regimes was also presented, alongside a pair of vacuum degenerate fields in arbitrary axisymmetric spacetimes. The construction of force-free fields in general axisymmetric geometry suggests that some observed jet features might not uniquely constrain the underlying spacetime. This degeneracy poses challenges for using field configuration to distinguish between black hole models.

These results offer a cohesive view of how electromagnetic fields evolve dynamically within different spacetime geometries. As the astrophysics community investigates alternatives to the Kerr solution for rotating black holes, this approach provides a tool for studying force-free fields in a broader range of spacetimes. Future research could focus on applying this formalism to more complex geometries, examining implications for astrophysical phenomena, and informing numerical simulations in extreme astrophysical environments. Possible search for Menon-Dermer-like fields in generalized Kerr spacetimes is relegated to the future.

\section*{Acknowledgements}
The author gratefully acknowledges insightful conversations with Bruno Costa and Govind Menon.

\printbibliography
\end{document}